\documentclass[aps,pre,amsmath,amssymb,twocolumn,groupedaddress,showpacs,eqsecnum]{revtex4}
\usepackage{graphicx}
\usepackage[utf8]{inputenc}
\usepackage[caption = false]{subfig}
\usepackage{float}

\usepackage{epsf}
\usepackage[toc]{appendix}
\newcommand{\be}{\begin{equation}}
\newcommand{\ee}{\end{equation}}
\newcommand{\ba}{\begin{eqnarray}}
\newcommand{\ea}{\end{eqnarray}}
\newcommand{\baa}{\begin{eqnarray}}
\newcommand{\eaa}{\end{eqnarray}}
\newcommand{\ed}{\end{document}}

\renewcommand{\baselinestretch}{1.2}
\date{\today}

\begin{document}

\title{Quantum gas in the fast forward scheme of adiabatically expanding cavities:\\ Force and equation of states}
\author{Gulmira Babajanova$^1$, Jasur Matrasulov$^1$ and Katsuhiro Nakamura$^{1,2}$}
\affiliation{$^{1}$Faculty of Physics, National University of Uzbekistan, Vuzgorodok, Tashkent 100174, Uzbekistan\\
$^{2}$Department of Applied Physics, Osaka City University, Sumiyoshi-ku, Osaka 558-8585, Japan
}

\begin{abstract} With use of the scheme of fast forward which realizes quasi-static or adiabatic dynamics in shortened time scale, we investigate a thermally-isolated ideal quantum gas confined in a rapidly dilating one-dimensional (1D) cavity with the time-dependent size $L=L(t)$. In the fast-forward variants of equation of states, i.e., Bernoulli's formula and Poisson's adiabatic equation, the force or 1D analog of pressure can be expressed as a function of the velocity ($\dot{L}$)  and acceleration ($\ddot{L}$)  of  $L$ 
besides rapidly-changing state variables like effective temperature ($T$) and $L$ itself.
The force is now a sum of nonadiabatic (NAD) and adiabatic contributions with the former caused by particles moving synchronously with kinetics of $L$ and the latter by ideal bulk particles insensitive to such a kinetics. The ratio of NAD and adiabatic contributions does not depend on the particle number ($N$)  in the case of the soft-wall confinement, whereas such a ratio is controllable in the case of hard-wall confinement. We also reveal the condition when the NAD contribution overwhelms the adiabatic one and thoroughly changes the standard form of the equilibrium equation of states.

\end{abstract}
\maketitle

\section{Introduction}

The equation of states plays an important role in thermodynamics and statistical mechanics. In constructing the equilibrium equation of states, the motion of the wall of a gas container (cylinder, cavity, billiard, {\it etc}.) is assumed to be quasi-static. In the Carnot's thermodynamic cycle \cite{carn,call,land}, the system undergoes very slowly  a series of different thermodynamic states and performs work on its surroundings. To make the theory of heat engines  realistic, however, one must evaluate the effect of a rapid wall motion of gas containers on the equation of states. In the context of a classical gas, Curzon and Ahlborn\cite{curz} and others\cite{vand,schm,izum,izum2}  investigated a finite-time heat engine. However, little attention has been paid to the nonequilibrium equation of states due to a rapidly-moving piston. 

In the case of Otto cycle undergoing alternately isentropic and isochore processes, the finite-time heat engine is being investigated for both single- and many-quantum particle systems in the case of the soft-wall confinement with a harmonic trap with time-dependent frequency \cite{aba1,aba2,aba3,zhen,beau}. But the researchers investigated neither  nonadiabatic force nor the nonequilibrium equation of states. 

In the equilibrium equation of states  for an ideal classical gas (Boltzmann gas), the pressure ($P$), volume ($V$) and temperature ($T$) are quasi-static state variables and satisfy Boyle-Charles' law (BCL) and Poisson's adiabatic equation (PAE) in the isothermal and thermally-adiabatic processes, respectively \cite{call,land}.  BCL is a special limit of the Bernoulli's formula (BF)  bridging between the pressure ($P$) and internal energy ($U$) for both quantum and classical gas in the cavity.  In the d-dimensional cavity, BF is given by $PV=\frac{2U}{d}$, which may be rewritten as $FL=2U$ with use of the force ($F$) and the length ($L$) in the case of one-dimensional (1D) cavity.  In the thermally-adiabatic process, PAE is given by $PV^{(d+2)/d}$=const.,  irrespective of classical and quantal systems. It  becomes $FL^3$=const. in 1D cavity.  

How will the above laws be innovated when the motion of a gas container would not be quasi-static?
While the perturbative analyses of a quantum gas \cite{avaz,sobi} were attempted in the linear response region, i.e., in the case of a piston with a small but finite velocity, the nonadiabatic contribution proved not to affect the equilibrium equation of states seriously. 
But, in the case of a rapid piston, we can expect a dramatic role of nonadiabatic contributions. 
The statistical treatment of a quantum gas is very difficult in general case of a rapid piston 
where the temporal change of state variables is far from being quasi-static. 
However, the fast forwarding of  adiabatic control \cite{adiab} of the confinement guarantees no transition among different quantum states, making such treatment feasible, and one can elucidate the exact relation among the rapidly-changing state variables.

Masuda and Nakamura \cite{mas1,mas2,mas3} proposed a way to
accelerate quantum dynamics with use of a characteristic driving
potential determined by the additional phase of a wave function.
See also a scheme for accelerating  quantum tunneling dynamics \cite{rv2-kn}.
This kind of acceleration
is called the fast forward, which means to reproduce a series of events or
a history of matters in a shortened time scale, like a rapid
projection of movie films on the screen. 

The fast forward theory applied to quantum adiabatic dynamics \cite{mas2,mas3,naka} needs no knowledge of spectral properties of the system and is free from the initial and boundary value problem. Therefore it constitutes one of the promising ways of shortcuts to adiabaticity (STA) devoted to tailor excitations in nonadiabatic processes\cite{dem1,dem2,ber1,lewi,chen,torr}.
Recent interesting application of the fast forward theory can be found in speedup of Dirac dynamics \cite{deff}, dynamical construction of classical adiabatic invariant \cite{jar1} and quasi-adiabatic spin dynamics of entangled states \cite{iwan}.
It is now timely to investigate the fast forward of the adiabatically-dilating cavities which contain the ideal quantum gas and to find its
nonequilibrium equation of states.


In this paper, confining ourselves to the thermally-isolated isentropic process where dynamics is unitary and the von Neumann entropy is constant, we shall investigate an ideal 1D quantum gas (Fermi gas) in the fast forward of the adiabatically-dilating cavities. Section \ref{sec2} is devoted to a brief summary of the latest variant \cite{naka} of the fast forward theory. 
In Section \ref{sec3}, we derive the fast-forward Hamiltonian for the harmonic oscillator with time-dependent frequency, to be used for the soft-wall confinement. In Section \ref{sec4}, we define the force operator due to a quantum gas in the case of rapidly-expanding or contracting cavities.  In Section \ref{sec5},  we shall solve the von Neumann equation, evaluate the statistical mean of the force operator, and obtain a fast-forward variants of  Bernoulli's formula and Poisson's adiabatic equation.
We study  the low-temperature quantal regime as well as the high-temperature quasi-classical regime and give physical interpretation of the results. Section \ref{sec6} is concerned with an analogous study in the case of the hard wall confinement. Summary and discussions are given in Section \ref{sec7}.

\section{Fast forward  of adiabatic dynamics}\label{sec2}

We shall sketch the scheme of fast forward of adiabatic control of 1 D confined
states.
Our strategy is as follows: (i) A given confining potential  $V_{0}$ is assumed to 
change adiabatically and to generate a stationary state $\psi_{0}$, which is an eigenstate
of the time-independent Schr\"{o}dinger equation with the instantaneous Hamiltonian.
Then both $\psi_{0}$ and $V_{0}$ are regularized so that they should satisfy the time-dependent Schr\"{o}dinger equation (TDSE); (ii) Taking the regularized state as a standard state, we shall change the time scaling with use of the scaling factor $\alpha\left(t\right)$, where the mean value $\bar{\alpha}$ of the infinitely-large time scaling factor $\alpha(t)$ will be chosen to compensate the infinitesimally-small growth rate $\epsilon$ of the quasi-adiabatic parameter and to satisfy $\bar{\alpha} \times \epsilon=finite$.

\subsection{Quasi-adiabatic dynamics}

Consider the standard dynamics with a deformable trapping potential whose shape is characterized by a slowly-varying control parameter $R(t)$ given by
\begin{eqnarray}
\label{2.1}\label{adia-R}
R(t)=R_{0}+\epsilon t,
\end{eqnarray}
with the growth rate $\epsilon\ll1$, which means that it requires a very long time $T=O\left(\frac{1}{\epsilon}\right)$, to see the recognizable change of $R(t)$. The time-dependent 1 D Schr\"{o}dinger equation (1D TDSE) for a charged particle is:
\begin{eqnarray}
\label{2.2}
\textit{i}\hbar\frac{\partial\psi_{0}}{\partial t}=-\frac{\hbar^{2}}{2m}\partial_x^{2}\psi_{0}+V_{0}(x,R(t))\psi_{0},
\end{eqnarray}
where the coupling with electromagnetic field is assumed to be absent.
The stationary bound state $\phi_{0}$ satisfies 
the time-independent counterpart given by
\begin{eqnarray}
\label{2.3}
E\phi_{0}=\hat{H}_0\phi_{0}\equiv\left[-\frac{\hbar^{2}}{2m}\partial_x^{2}+V_{0}(x,R)\right]\phi_{0}.
\end{eqnarray}

Then, with use of the eigenstate $\phi_0=\phi_0(x, R)$ satisfying Eq.(\ref{2.3}),
one might conceive the corresponding time-dependent state to be a product of $\phi_0$ and a dynamical factor as,
\begin{eqnarray}\label{eq3.4}
\psi_{0}=\phi_{0}(x,R(t))e^{-\frac{\textit{i}}{\hbar}\int^{t}_{0}E(R(t'))dt'}.
\end{eqnarray}
As it stands, however, $\psi_{0}$ does not satisfy TDSE in Eq.(\ref{2.2}). Therefore we introduce a regularized state
\begin{eqnarray}\label{eq3.5}
\psi^{reg}_{0}
&\equiv& \phi_{0}(x,R(t))e^{\textit{i}\epsilon\theta(x,R(t))}e^{-\frac{\textit{i}}{\hbar}\int^{t}_{0}E(R(t'))dt'}
\nonumber\\
&\equiv& \phi_0^{reg} (x,R(t))e^{-\frac{\textit{i}}{\hbar}\int^{t}_{0}E(R(t'))dt'}
\end{eqnarray}
together with a regularized potential
\begin{eqnarray}\label{eq3.6}
V^{reg}_{0}\equiv V_{0}(x,R(t))+\epsilon\tilde{V}(x,R(t)).
\end{eqnarray}
The unknown $\theta$ and $\tilde{V}$ will be determined self-consistently so that $\psi^{reg}_{0}$ should fulfill the TDSE,
\begin{eqnarray}\label{eq3.7}
\textit{i}\hbar\frac{\partial\psi_{0}^{reg}}{\partial t}=-\frac{\hbar^{2}}{2m}\partial_x^{2}\psi^{reg}_{0}+V^{reg}_{0}\psi^{reg}_{0},
\end{eqnarray}
up to the order of $\epsilon$.

Rewriting $\phi_{0}(x,R(t))$ with use of the real positive amplitude $\overline{\phi}_{0}(x,R(t))$ and phase $\eta(x,R(t))$ as
\begin{eqnarray}\label{2.8}
\phi_{0}(x,R(t))=\bar{\phi}_{0}(x,R(t))e^{\textit{i}\eta(x,R(t))},
\end{eqnarray}
we see $\theta$ and $\widetilde{V}$ to satisfy:
\begin{eqnarray}\label{2.9}
\partial_x(\bar{\phi}_{0}^2\partial_x\theta) =-\frac{m}{\hbar}\partial_{R}\bar{\phi}_{0}^2,
\end{eqnarray}
\begin{eqnarray}\label{2.10}
\frac{\tilde{V}}{\hbar}=-\partial_{R}\eta-\frac{\hbar}{m}\partial_x\eta\cdot\partial_x\theta.
\end{eqnarray}
Integrating Eq. (\ref{2.9}) over  $x$, we have
\begin{equation}\label{2.11}
\partial_x\theta=-\frac{m}{\hbar}\frac{1}{\bar{\phi}_0^2}\int^x \partial_R\bar{\phi}_0^2 dx', 
\end{equation}
which is the core equation of the regularization procedure. The problem of singularity due to nodes of $\bar{\phi}_{0}$ in Eq.(\ref{2.11}) can be overcome, so long as one is concerned with the systems with scale-invariant potentials 
(see Section \ref{sec3}).
 
\subsection{Exact fast-forwarding with use of magnified time scale} 

We shall now accelerate the quasi-adiabatic dynamics of $\psi_{0}^{reg}$ in Eq.(\ref{eq3.5}), by applying the electromagnetic field.

We first introduce the fast-forward version 
of $\psi_{0}^{reg}$ as
\begin{eqnarray}\label{2.12}
\psi_{FF}^{(0)}(x,t)&\equiv& \psi^{reg}_{0}(x,R(\Lambda(t)))\nonumber\\
&\equiv& \phi^{reg}_{0}(x,R(\Lambda(t)))e^{-\frac{\textit{i}}{\hbar}\int^{t}_{0}E(R(\Lambda(t')))dt'}\nonumber\\
\end{eqnarray}
with
\begin{equation}\label{rv2-1}
R(\Lambda(t))=R_0+\epsilon \Lambda(t),
\end{equation}
where $\Lambda(t)$ is the future or advanced time 
\begin{eqnarray}\label{1.3}
\Lambda(t)=\int_0^t\mathrm{\alpha(t')}\,\mathrm{d}t',
\end{eqnarray}
and $\alpha (t)$ is a magnification scale factor defined by
$\alpha(0)=1$, $\alpha(t)>1$  $(0\le t \le T_{FF})$,
$\alpha(t)=1$ $(t  > T_{FF})$.
Suppose $T$ to be a very long time to see a recognizable change of the adiabatic parameter $R(t)$ in Eq.(\ref{adia-R}), and then the corresponding change of $R(\Lambda(t))$ is realized in the shortened or fast-forward time $T_{FF}$ defined by
\begin{eqnarray}\label{T-Tff}
T=\int_0^{T_{FF}}\alpha (t) \mathrm{d}t.
\end{eqnarray}

The explicit expression for $\alpha(t)$ in the fast-forward range ($0 \le t \le T_{FF}$) is:
\begin{eqnarray} 
\label{1.7}
\alpha(t)=\bar{\alpha}- (\bar{\alpha}-1)\cos (\frac{2\pi}{T/\bar{\alpha}}t),
\end{eqnarray}
where $\bar{\alpha}$ is the mean value of $\alpha(t)$ and is given by $\bar{\alpha}=T/T_{FF}$ \cite{mas1,mas2,mas3}.

Then let's assume $\psi_{FF}^{(0)}$ to be the solution of the TDSE  for a charged particle in the presence of  gauge potentials, 
$A_{FF}^{(0)}(x,t)$ and 
$V_{FF}^{(0)}(x,t)$,
\begin{eqnarray}
\label{1.9}
&&\imath\hbar\frac{\partial{\psi_{FF}^{(0)}}}{\partial{t}}=H_{FF}\psi_{FF}^{(0)}\equiv \nonumber\\
&&\left(\frac{1}{2m}(\frac{\hbar}{i}\partial_x - A_{FF}^{(0)})^2+ V_{FF}^{(0)}+V_0^{reg}\right)\psi_{FF}^{(0)},\nonumber\\
\end{eqnarray}
where, for simplicity, we employ the prescription of a positive unit charge ($q=1$) and the unit velocity of light ($c=1$). The driving electric field is given by,
\begin{eqnarray}
\label{1.10}
E_{FF}=-\frac{\partial A_{FF}^{(0)}}{\partial t}-\partial_x V_{FF}^{(0)}.
\end{eqnarray}

Substituting Eq.(\ref{2.12}) into Eq.(\ref{1.9}), we find
$\phi^{reg}_{0}$ to satisfy
\begin{eqnarray}
\label{2.13}
\textit{i}\hbar\frac{\partial\phi^{reg}_{0}}{\partial t}&=&\frac{1}{2m}\left(\frac{\hbar}{\textit{i}}\partial_x-A_{FF}^{(0)}\right)^{2}\phi^{reg}_{0} \nonumber\\
&+&(V_{FF}^{(0)}+V_0^{reg}-E)\phi^{reg}_{0},
\end{eqnarray}
where $V_0^{reg}\equiv V^{reg}(x,R(\Lambda(t)))$, i.e., the advanced-time variant of Eq.(\ref{eq3.6}).
The dynamical phase in Eq.(\ref{2.12}) has led to
the energy shift in the potential in Eq.(\ref{2.13}).

Rewriting $\phi^{reg}_{0}$ in terms of the amplitude $\bar{\phi}_{0}$ and phases $\eta+\epsilon\theta$ as
\begin{eqnarray}\label{eq3.13}
\phi^{reg}_{0}\equiv  \bar{\phi}_{0}(x,R(\Lambda(t)))e^{\textit{i}\left[\eta(x,R(\Lambda(t)))
+\epsilon\theta(x,R(\Lambda(t)))\right]}, \nonumber\\
\end{eqnarray}
and using Eq.(\ref{eq3.13}) in Eq.(\ref{2.13}), we find $A_{FF}^{(0)}$ of $O(\epsilon\alpha)$ and $V_{FF}^{(0)}$ consisting of terms of $O(\epsilon\alpha)$ and $O((\epsilon\alpha)^2)$.

Now, applying our central strategy to take the limit $\epsilon\rightarrow 0$ and $\bar{\alpha}\rightarrow \infty$ with $\epsilon\bar{\alpha}=\bar{v}$ being kept finite, we can reach the issue (for details, see \cite{naka}):
\begin{eqnarray}
\label{2.20}
A_{FF}^{(0)}&=&-\hbar v(t)\partial_x\theta,\nonumber\\
V_{FF}^{(0)}&=&-\frac{\hbar^{2}}{m}v(t)\partial_x\theta\cdot\partial_x\eta \nonumber\\
&-&\frac{\hbar^{2}}{2m}(v(t))^{2}(\partial_x\theta)^{2}-\hbar v(t)\partial_{R}\eta,
\end{eqnarray}
where, with use of $T_{FF}\left(=\frac{T}{\bar{\alpha}}=O\left(\frac{1}{\epsilon\bar{\alpha}}\right)\right)=finite$,
%
\begin{eqnarray}
\label{2.21}
v(t)&\equiv&\lim_{\epsilon\rightarrow 0, \bar{\alpha}\rightarrow \infty}\varepsilon\alpha(t)=\bar{v}\left(1-\cos\frac{2\pi}{T_{FF}}t\right),\nonumber\\
R(\Lambda(t))&=&R_0+\lim_{\epsilon\rightarrow 0, \bar{\alpha}\rightarrow \infty}\varepsilon\Lambda(t)\nonumber\\
&=&R_{0}+\int^{t}_{0}v(t')dt'\nonumber\\
&=& R_{0}+\bar{v}\left(t-\frac{T_{FF}}{2\pi}\sin\left(\frac{2\pi}{T_{FF}}t\right)\right),\nonumber\\
&& \;{\rm for} \; 0  \le t \le T_{FF},
\end{eqnarray}
and
\begin{eqnarray}\label{beyon}
v(t)=0,  \quad R(\Lambda(t))=R_0+\bar{v}T_{FF}  \quad {\rm for} \;  t >T_{FF} .\nonumber\\
\end{eqnarray}
$v(t)$ and its mean $\bar{v}$ stand for the time-scaling factors coming from $\alpha(t)$ and $\bar{\alpha}$, respectively.

In the same limiting case as above, $\psi_{FF}^{(0)}$ is explicitly given by
\begin{eqnarray}
\label{2.23}
\psi_{FF}^{(0)}=\bar{\phi}_{0}(x,R(\Lambda(t)))
e^{\textit{i}\eta(x,R(\Lambda(t)))}e^{-\frac{\textit{i}}{\hbar}\int^{t}_{0}E(R(\Lambda(t')))dt'}. \nonumber\\
\end{eqnarray}

\subsection{Gauge transformation and  fast-forwarding with the extra phase factor} 

While the scheme so far guarantees the fast forward of both the amplitude and phase of wave functions, it is now convenient to construct a $A_{FF}$-free variant of the scheme. Let us introduce the gauge transformation of Eqs.  (\ref{2.20}), and (\ref{2.23}) as follows
\begin{eqnarray}\label{eq3.20}
\psi_{FF}^{(0)}&\rightarrow& \psi_{FF}e^{\textit{-if}},\nonumber\\
V_{FF}^{(0)}&\rightarrow& V_{FF}+\hbar\partial_{t}\textit{f},\nonumber\\
A_{FF}^{(0)}&\rightarrow& A_{FF}-\hbar\partial_x\textit{f},
\end{eqnarray}
where the phase $f$ defined so as to cancel $A_{FF}^{(0)}$ in Eq.(\ref{2.20}) and to make $A_{FF}=0$  is given by
\begin{eqnarray}\label{eq3.21}
f=v(t)\theta(x,R(\Lambda(t))).
\end{eqnarray}
$\theta$ and $v(t)$ are available from Eq.(\ref{2.11}) and Eq.(\ref{2.21}), respectively.
This gauge transformation leads to the fast-forward state
with the extra phase as:
\begin{eqnarray}
\label{3.13}
\psi_{FF}&=\bar{\phi}_{0}(x,R(\Lambda(t)))
e^{\textit{i}\eta(x,R(\Lambda(t)))}e^{\textit{i}v(t)\theta(x,R(\Lambda(t)))}\nonumber\\
&\times e^{-\frac{i}{\hbar}\int_0^t E(R(\Lambda(s)))ds},
\end{eqnarray}
which satisfies TDSE with a fast-forward Hamiltonian $H_{FF}$:
\begin{eqnarray}
\label{3.15}
\textit{i}\hbar\frac{\partial \psi_{FF}}{\partial t}=H_{FF} \psi_{FF}\equiv \left(-\frac{\hbar^{2}}{2m}\partial_x^2+V_{0}
+ V_{FF}\right)\psi_{FF}.\nonumber\\
\end{eqnarray}
Here $V_0=V_0(x, R(\Lambda(t)))$ and $V_{FF}$ is given by
\begin{eqnarray}
\label{3.14}
V_{FF}&=&-\frac{\hbar^{2}}{m}v(t)\partial_x\theta\cdot\partial_x\eta-\frac{\hbar^{2}}{2m}(v(t))^{2}(\partial_x\theta)^{2}\nonumber\\
&&-\hbar v(t)\partial_{R}\eta-\hbar\dot{v}(t)\theta-\hbar(v(t))^{2}\partial_{R}\theta.\nonumber\\
\end{eqnarray}

Equations (\ref{3.13}),(\ref{3.15}) and (\ref{3.14}) are the issue of the fast forward theory. $V_{FF}$ is responsible for the driving electric field
$E_{FF}=-\partial_xV_{FF}$, and guarantees the fast-forward state in Eq.(\ref{3.13}).
For our study below, it is convenient to rewrite the above issue with use of quantum numbers.
Let's rewrite eigenstates and eigenvalues of Eq.(\ref{2.3}) as $\left|n(R)\right\rangle^{(0)}$ and $E_n(R)$, respectively.
Then $\psi_{FF}$ in Eq.(\ref{3.13}) is expressed as $\left|n\right\rangle\equiv \left|n(R(\Lambda(t)))\right\rangle\equiv \left|n(R(\Lambda(t)))\right\rangle^{(0)} e^{iv(t)\theta(x,R(\Lambda(t)))}\cdot e^{-\frac{i}{\hbar}\int_0^t E_n(R(\Lambda(s)))ds}$, 
and TDSE in Eq.(\ref{3.15}) as 
$i\hbar\left|\dot{n}\right\rangle=H_{FF}\left|n\right\rangle$. Finally, $\{\left|n\right\rangle\}$ satisfies the completeness condition
$\sum_{n=0}^{\infty}\left|n\rangle \langle n\right|=\sum_{n=0}^{\infty}\left|n\rangle^{(0)(0)}\langle n\right|=$I. These $n$-dependent
variant of the issue of the fast forward theory will be repeatedly used in the following Sections.

\section{Acceleration of adiabatic control of soft-wall confinement}\label{sec3}

In this Section, we shall investigate the harmonic oscillator with time-dependent frequency and obtain its fast-forward Hamiltonian  $H_{FF}$ to be used in the statistical treatment of a confined quantum gas.

\subsection{Scale-invariant bound systems in the context of fast forwarding}\label{scale-FF}
Consider the original potential controlled by the scale-invariant adiabatic 
expansion and contraction \cite{ber2,cam1,def2}, as given by
\begin{eqnarray}
\label{eqa5}
V_0=\frac{1}{R^2}U_0\left(\frac{x}{R}\right),
\end{eqnarray}
where $R$ is the adiabatic parameter as in Eq.(\ref{2.1}). The corresponding 1 D eigenvalue problem for bound systems yields ground and excited states whose normalized forms are commonly given by
\begin{eqnarray}
\label{3.17}
\phi_0=\frac{1}{\sqrt{R}}h \left(\frac{x}{R}\right),
\end{eqnarray}
where $h=\bar{h}e^{i\eta}$ with real amplitude $\bar{h}$ and phase $\eta$.
Then, with use of a new variable $X\equiv \frac{x}{R}$, Eq.(\ref{2.11}) becomes
\begin{eqnarray}
\label{3.18}
\partial_x \theta=-\frac{m}{\hbar} \frac{R}{|\bar{h}(X)|^2} 
\partial_R \int^X |\bar{h}(X^{\prime})|^2 d X^{\prime}.
\end{eqnarray}
Here the indefinite integral is used because the lower limit of integration is arbitrary. Noting $\partial_R=\frac{\partial X }{\partial R} \frac{\partial }{\partial X}=-\frac{x}{R^2} \frac{\partial }{\partial X}$, Eq.(\ref{3.18}) reduces to
\begin{eqnarray}
\label{3.19}
\partial_x \theta= \frac{m}{\hbar} \frac{x}{R} \frac{|\bar{h}(X)|^2}{|\bar{h}(X)|^2}=\frac{m}{\hbar R} x.
\end{eqnarray}
In the second equality of Eq.(\ref{3.19}), we prescribed $\lim_{X\rightarrow X_c}\frac{|\bar{h}(X)|^2}{|\bar{h}(X)|^2}=1$ if $\bar{h}(X)$ will be $\bar{h}(X_c)=0$ at $X=X_c$. From Eq.(\ref{3.19}), one finds:
\begin{eqnarray}\label{eqa9}
\theta&=&\frac{m}{2\hbar R} x^2,\nonumber\\
\partial_R \theta&=&-\frac{m}{2\hbar R^2} x^2.
\end{eqnarray}
In the simple case that $\phi_0$ in Eq.(\ref{3.17}) is real, i.e., $\eta=0$,  
$V_{FF}$ in Eq. (\ref{3.14}) becomes
\begin{eqnarray}
\label{3.21}
V_{FF}=-\frac{m\ddot{R}}{2R}x^2,
\end{eqnarray}
where $R=R(\Lambda(t))$, $v(t)=\dot{R}$ and $\dot{v}(t)=\ddot{R}$ in Eq.(\ref{2.21}) are used. $V_{FF}$ in Eq.(\ref{3.21}) is nothing but the counter-diabatic potential in the scale-invariant bound systems \cite{cam1,def2}. The electric field is now given by
\begin{eqnarray}
E_{FF}=-\frac{\partial}{\partial x}V_{FF}=\frac{m\ddot{R}}{R}x.
\end{eqnarray}

Thus the fast forward approach applied to the scale-invariant bound systems is free from the problem of singularity 
caused by nodes of eigenstates.

\subsection{Harmonic oscillator with time-dependent frequency}

Let us investigate a quantum harmonic oscillator with time-dependent frequency, which constitutes a special bound system with the scale-invariant potential.
The original adiabatic dynamics is described by 
\begin{eqnarray}\label{leftWF}
\textit{i}\hbar\frac{\partial}{\partial t}\psi_{0}(x,R(t))&=& H_{0}(x,R(t))\psi_{0}(x,R(t))
\end{eqnarray}
with
\begin{eqnarray}\label{rightWF}
H_{0}(x,R(t))=-\frac{\hbar^{2}}{2m}\partial_x^{2}+\frac{1}{2}m\omega^{2}(R(t))x^{2}.
\end{eqnarray}
Here $ \omega=\omega\left(R\left(t\right)\right) $ is the frequency which varies slowly through the adiabatic parameter $R$ in Eq.(\ref{adia-R}).
Comparing $V_{0}=\frac{1}{2}m\omega^2 x^2$ in Eq.(\ref{rightWF}) with scale-invariant expression in Eq.(\ref{eqa5}), we see
 \begin{eqnarray}
 R\left(t\right)=\frac{1}{\sqrt{\omega}}.
 \end{eqnarray}

On the other hand, the effective size $L(t)$ of the wave function is obtained from the adiabatic eigenvalue problem 
for the Hamiltonian in Eq.(\ref{rightWF}) and is given by
 \begin{eqnarray}\label{lt}
 L\left(t\right)=\sqrt{\frac{\hbar}{m\omega\left(t\right)}}.
 \end{eqnarray}
 Therefore $R(t)$ is now read as $L(t)$ up to the multiplication factor $\sqrt{\frac{\hbar}{m}}$.
 
 The adiabatic eigenvalue problem
\begin{eqnarray}\label{H-E}
H_{0}(x,L)\phi=E(L)\phi
\end{eqnarray}
gives the eigenvalue and eigenstate as 
\begin{eqnarray}\label{3.5}
E_{n}&=&\left( n+\frac{1}{2}\right) \hbar\omega(L),\nonumber\\
\phi_{n}&=&\left( \frac{m\omega(L)}{\pi\hbar}\right) ^\frac{1}{4}\frac{1}{(2^n n!) ^\frac{1}{2}}\nonumber\\
&\times& e^{-\frac{m\omega(L)}{2\hbar}x^{2}} H_{n}\left(\left(\frac{m\omega(L)}{\hbar} \right)^\frac{1}{2}x \right)
\end{eqnarray}
with $n=0,1,2,\cdots$ . Here  $ H_{n}(\cdot)$s are Hermite polynomials.
 
Applying the result in Eq.(\ref{3.13}) together with Eq.(\ref{eqa9}) and $v=\dot{L}$, the fast forward state is given by
\begin{eqnarray}\label{Psi-F}
\psi_{FF}&=&\phi_{n}(x,L(\Lambda(t))e^{i\frac{m}{2\hbar}\frac{\dot{L}}{L}x^2} e^{-\left( n+\frac{1}{2}\right) i \int_{0}^{t}\omega(L(\Lambda(t')))dt'}\nonumber\\
&\equiv & <x|n>,
\end{eqnarray}
which satisfies TDSE in Eq.(\ref{3.15}). The fast forward Hamiltonian becomes:
\begin{eqnarray}\label{HFF-HO}
H_{FF}=\frac{p^2}{2m}+V_{0}+V_{FF}
\end{eqnarray}
with
\begin{eqnarray}\label{VFF-HO}
V_{0}+V_{FF}=\frac{1}{2}m\left(\frac{\hbar^2}{m^2}\frac{1}{L^4}-\frac{\ddot{L}}{L}\right)x^2,
\end{eqnarray} 
where the general result in Eq.(\ref{3.21}) is used.

The effective confining size $L$ and the frequency $\omega$ are now expressed as
\begin{eqnarray}\label{FF-paramet}
L\equiv L(\Lambda(t))&=&L_0 + \int_0^t v(t')dt', \nonumber\\
\frac{\omega(L(\Lambda(t)))}{\omega_0}&\equiv & \left( \frac{L_0}{L(\Lambda(t))}\right)^2.
\end{eqnarray}

\section{Force operator}\label{sec4}

We now embark upon the statistical treatment of a quantum gas (Fermi gas) of non-interacting particles confined in a harmonic potential with its frequency being time-dependent. When the confined region will be increased or expanded, the gas system exerts a force on its outside. The force operator and its statistical mean play an essential role in constructing the equation of states. The force consists of both adiabatic and nonadiabatic parts, when the temporal change of the confining area is not quasi-static.




For the soft-walled confinement with a time-dependent effective size $L=L(\Lambda(t))$ in Eq.(\ref{FF-paramet}),
the fast-forward Hamiltonian $H_{FF}$ is explicitly given by Eqs.(\ref{HFF-HO}) and (\ref{VFF-HO}).
We now see the expectation of $H_{FF}$ as given by
\begin{equation}
\langle \psi_{FF} |H_{FF}|\psi_{FF} \rangle,
\end{equation}
where $|\psi_{FF}  \rangle$ is a solution of TDSE in Eq.(\ref{3.15}) with Hamiltonian $H_{FF}$.

 The expectation of the force acting on the wall is obtained by
\begin{equation}
\left\langle F\right\rangle=-\frac{\partial}{\partial L} \langle \psi_{FF} |H_{FF}|\psi_{FF}\rangle.
\label{4.4}
\end{equation}
Noting $\frac{\partial}{\partial L} |\psi_{FF} \rangle=\frac{1}{\dot{L}} \frac{\partial}{\partial t} |\psi_{FF} \rangle=\frac{1}{i\hbar \dot{L}}H_{FF}|\psi_{FF} \rangle$ and its Hermitian conjugate, Eq. (\ref{4.4}) reduces to
\begin{equation}
\left\langle F\right\rangle=-\langle \psi_{FF} |\frac{\partial H_{FF}}{\partial L}|\psi_{FF} \rangle.
\label{averF2}
\end{equation}
Hence the force operator is defined by
\begin{equation}
\hat{F}= -\frac{\partial H_{FF}}{\partial L}.
\label{opeF}
\end{equation}

However, the kinetic energy of $H_{FF}$ does not include $L$ explicitly. Therefore it is not obvious how to evaluate the force operator directly by using Eq.(\ref{opeF}).

To overcome this difficulty, we shall first make  the time-dependent canonical transformation related to the scale transformation of both the coordinate $x$ and amplitude of wave function as
\begin{eqnarray}
H_{\Gamma}=e^{-iU}(H_{FF}-i\hbar \frac{\partial}{\partial t} )e^{iU},
\end{eqnarray}
where
\begin{eqnarray}
U=-\frac{1}{2\hbar}(\hat{x}\hat{p}&+&\hat{p}\hat{x})\ln L=i\left(x\frac{\partial}{\partial x}+\frac{1}{2} \right)\ln L.\nonumber\\
\end{eqnarray}
Similarly the amplitude of the wave function is scaled as
\begin{eqnarray}
\phi_{\Gamma}= e^{-iU}\psi_{FF}(x,t).
\end{eqnarray}
Finally the Schr\"odinger equation is transformed to
\begin{eqnarray}
i\hbar \frac{\partial }{\partial t}\phi_{\Gamma}=H_{\Gamma}\phi_{\Gamma} 
\label{4.11}
\end{eqnarray}
with the new Hamiltonian
\begin{eqnarray}
H_{\Gamma}= -\frac{\hbar^2}{2m}\frac{1}{L^2}\frac{\partial^2 }{\partial x^2}+
i\hbar\frac{\dot{L}}{L}x\frac{\partial }{\partial x}+\nonumber\\
\frac{i\hbar}{2} \frac{\dot{L}}{L}+\frac{1}{2}m\left(\frac{\hbar^2}{m^2}\frac{1}{L^2}-\ddot{L}L\right)x^2.
\label{4.12}
\end{eqnarray}
Taking $L$ derivative of $H_{\Gamma}$, we can rigorously define the force operator in the transformed space as
\begin{eqnarray}\label{pseudo-f}
F_{\Gamma}&=&-\frac{\partial H_{\Gamma}}{\partial L}\nonumber\\
&=& \frac{1}{mL^3}\tilde{p}_x^2-\frac{\dot{L}}{2L^2}\left(\tilde{x}\tilde{p}+\tilde{p}\tilde{x} \right) +\frac{1}{2}m\left(\frac{\hbar^2}{m^2}\frac{2}{L^3}+\ddot{L}\right)x^2.\nonumber\\
\end{eqnarray}

Now, carrying out the inverse canonical transformation ($xL \to x$, etc.),
we have the force operator expressed in the original space as
\begin{eqnarray}
\label{4.15}
\hat{F}&=&e^{iU} F_{\Gamma} e^{-iU}=\nonumber\\
&=&\frac{\hat{p}^2}{mL} - \frac{\dot{L}}{2L^2} (\hat{x}\hat{p} +\hat{p}\hat{x})+
\frac{1}{2}m\left(\frac{\hbar^2}{m^2}\frac{2}{L^5}+\frac{\ddot{L}}{L^2}\right)\hat{x}^2,\nonumber\\
\end{eqnarray}
which certainly satisfies:
\begin{eqnarray}
\langle \psi_{FF} |\hat{F}|\psi_{FF} \rangle \equiv -\frac{\partial}{\partial L}\langle \psi_{FF} |H_{FF}|\psi_{FF} \rangle,
\end{eqnarray}
where $\psi_{FF}$ is given in Eq.(\ref{Psi-F}).
In fact, 
\begin{eqnarray}\label{force-exp}
\langle \psi_{FF} |\hat{F}|\psi_{FF} \rangle&\equiv&\langle n|\hat{F}|n\rangle \nonumber\\
&=&\frac{\hbar^2}{m}\frac{2n+1}{L^3} + \frac{2n+1}{4}m\ddot{L}\equiv F_n \nonumber\\
\end{eqnarray}
is available from the variational derivative of
\begin{eqnarray}\label{hamil-exp}
&&\langle \psi_{FF} |H_{FF}|\psi_{FF} \rangle \equiv \langle n|H_{FF}|n\rangle \nonumber\\
&=&\frac{2n+1}{2}\frac{\hbar^2}{m}\frac{1}{L^2}+\left(n+\frac{1}{2}\right)\frac{m}{2}\dot{L}^2-\left(n+\frac{1}{2}\right)\frac{m}{2}\ddot{L}L  \nonumber\\
\end{eqnarray}
with respect to $L$.
In Eq. (\ref{4.15}), the adiabatic force corresponds to the terms depending only $L$. The nonadiabatic force corresponds to those dependent on $\dot{L}$ and $\ddot{L}$, and are time-reversal symmetric, namely, invariant against the operation $\dot{L}\rightarrow -\dot{L}, \ddot{L} \rightarrow \ddot{L}, \hat{p} \rightarrow -\hat{p}$.

\begin{figure}[H]
\begin{center}
\includegraphics[width=3in]{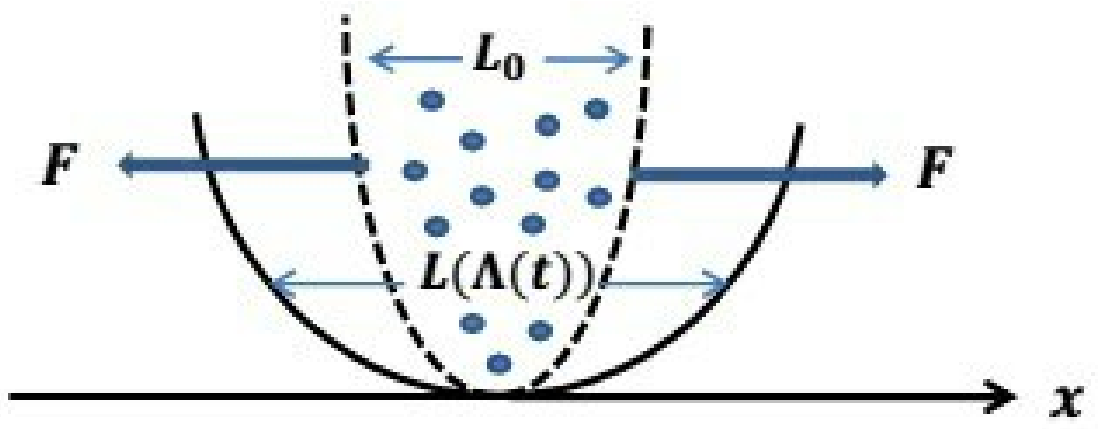}
\includegraphics[width=3in]{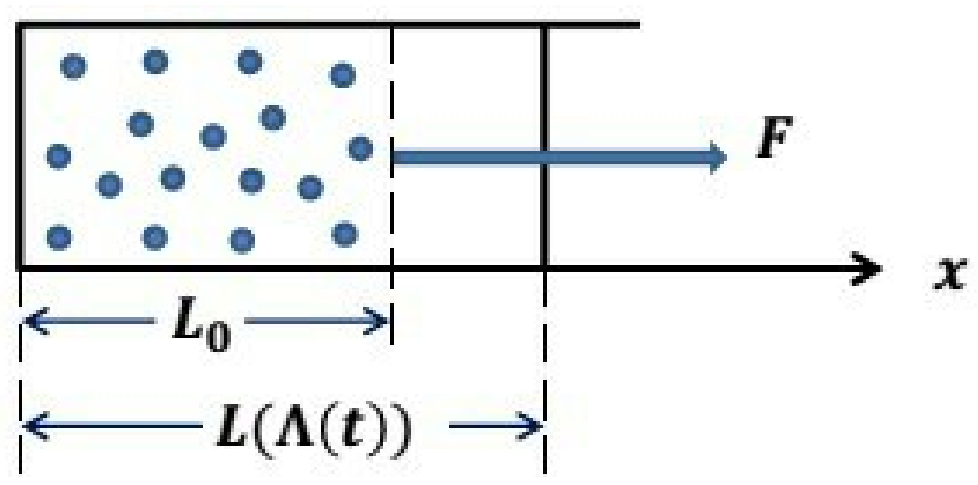}
\caption{Upper panel: quantum gas in soft-wall confinement; Lower panel: quantum gas in hard wall confinement. $L_0$ and $L(\Lambda(t))$ are the initial and time-dependent size of the expanding cavities, respectively. $F$ is the force due to the gas.} 
\end{center}
\end{figure}

\section{Nonequilibrium equation of state}\label{sec5}

Let's enter to the main part of the present paper. In this Section we consider a Fermi gas of $N$ noninteracting particles confined in harmonic potential whose frequency is time-dependent and in the next Section we shall move to the gas system in the hard-wall confinement. Figure 1 illustrates two kind of confinements. We shall derive the nonequilibrium equation of states during the fast forward.
The statistical mean of the force operator is given by 
\begin{eqnarray}\label{st-mean-F}
\bar{F}=\rm{Tr}\left(\rho\hat{F}\right). 
\end{eqnarray}
Here $\rho$ is the density operator for the mixed state satisfying the von Neumann equation
\begin{eqnarray}
i\hbar\frac{\partial\rho}{\partial t}=\left[H_{FF},\rho\right].
\end{eqnarray}
With use of the exact solution $\left\{\left|n\right\rangle\right\}$ of TDSE in Eq.(\ref{3.15}) and noting the quantum-number dependent description of the fast forward theory in the last paragraph of Section \ref{sec2}, $\rho$ is solved as 
\begin{eqnarray}\label{rho-expand}
\rho\left(t\right)=\sum^{\infty}_{n=0}\left|n\right\rangle f_{n}\left\langle n\right|,\nonumber\\
\end{eqnarray}
where $f_{n}$ is the Fermi-Dirac distribution at the initial time ($t=0$), which, takes
\begin{eqnarray}\label{FD-init}
f_{n}=\frac{1}{e^{\beta\left(E_{n}(L(\Lambda(t=0)))-\mu\right)}+1}\equiv\rho_{nn}\left(0\right).\nonumber\\
\end{eqnarray}

In fact, with use of TDSE $i\hbar\left|\dot{n}\right\rangle=H_{FF}\left|n\right\rangle$, we see
\begin{eqnarray}
i\hbar \dot{\rho}&=&\sum_{n}\Bigl(i\hbar\left|\dot{n}\right\rangle f_n\left\langle n\right|-\left|n\right\rangle f_n \left\langle \dot{n}\right|i\hbar\Bigr) \nonumber\\
&=&H_{FF}\sum_{n}\left|n\right\rangle f_n \left\langle n\right|-\sum_{n}\left|n\right\rangle f_n \left\langle n\right| H_{FF} \nonumber\\
&\equiv &\left[H_{FF},\rho\right].
\end{eqnarray} 
   

\subsection{Low-temperature ($T \ll T_{0}$) and high-density region }
At $T=0$ (zero temperature) or $\beta\left(\equiv\frac{1}{k_{B}T}\right)=+\infty$, $f_{n}$ reduces to the Heaviside step function $f_{n}=\Theta\left(E_{F}-E_{n}\right)$.
Then 
\begin{eqnarray}
\bar F=2\sum^{\infty}_{n=0}f_{n}F_{n}\equiv\bar F^{ad}+\bar F^{nad},
\end{eqnarray}
where, with use of Eq.(\ref{force-exp}), 
\begin{eqnarray}
\bar F^{ad}&=&2\sum^{N+\frac{1}{2}}_{n=0}\frac{\hbar^2}{m}\frac{2n+1}{L^3}=\frac{\hbar^2}{m}\frac{N^2}{2L^3},\nonumber\\
\bar F^{nad}&=&2\sum^{N+\frac{1}{2}}_{n=0}\frac{2n+1}{4}m\ddot{L}=\frac{m}{8}N^2\ddot{L} 
\end{eqnarray}
for  the total number of electrons $N\gg 1$.
As we shall see below, $\bar F^{nad}$ plays a role in the nonequilibrium equation of states in the fast-forward protocol of a very rapid piston. 

At $T\neq0$ (finite temperature) or $\beta\left(\equiv\frac{1}{k_{B}T}\right)<+\infty$, we firstly summarize the formula for thermodynamic averages, with use of the Fermi-Dirac distribution $f(E)=\frac{1}{e^{\beta(E-\mu)}+1}$ .

In low-temperature region at $T\ll T_0$ (degenerate temperature), we see the formula  \cite{land}:
\begin{eqnarray}
\label{4.25}
\int_{E_0}^\infty g(E)f(E)dE&=&\nonumber\\
=\int_{E_0}^\mu g(E)dE&+&\frac{\pi^2(kT)^2}{6} g'(\mu)+O((kT)^4),\nonumber\\
\end{eqnarray} 
where $g(E_0)=0$ is assumed.

Choosing 1 D density of states $D(E)$ as $g(E)$, we have
\begin{eqnarray}
\label{4.27}
N=2\int_{0}^\infty D(E)f(E)dE,
\end{eqnarray}
which defines the chemical potential $\mu$ as a function $N$.

At $t=0$, the adiabatic eigenvalues are
\begin{eqnarray}\label{energ-init}
E_{n}=\left(n+\frac{1}{2}\right)\hbar\omega_0, \left(n=0,1,2,\cdots\right), 
\end{eqnarray}
from which we can obtain density of states as
\begin{eqnarray}\label{dens-init}
\lim_{\Delta E\rightarrow 0}\frac{\Delta n}{\Delta E}\equiv D\left(E\right)=\frac{1}{\hbar\omega_0}.
\end{eqnarray}
With use of Eq.(\ref{4.27}) and $\hbar\omega_0=\frac{\hbar^2}{mL_{0}^2}$ (see Eq.(\ref{FF-paramet})), the chemical potential is obtained as
\begin{eqnarray}\label{1d-lt-che}
\mu=\frac{N}{2}\frac{\hbar^2}{mL_{0}^2}.
\end{eqnarray} 
Having recourse to formulas in Eqs. (\ref{force-exp}), (\ref{FD-init}), (\ref{energ-init}), (\ref{dens-init}) and (\ref{1d-lt-che}),  the expectation of force becomes
\begin{eqnarray}
\label{force}
\bar{F}&=&2\sum^{\infty}_{n=0}f_{n}F_{n}=N^2\left(\frac{\hbar^2}{2mL^3}+\frac{m}{8} \ddot{L}\right)\nonumber\\
&\times&\left[1+\frac{4\pi^2}{3}L_{0}^2\left(\frac{mkT}{\hbar^2}\right)^2 \left(\frac{N}{L_{0}}\right)^{-2}+\cdots\right],\nonumber\\
\end{eqnarray}
the leading term of which indicates that the force consists of  
adiabatic ($\frac{\hbar^2}{2mL^3}$) and nonadiabatic ($\frac{m}{8} \ddot{L}$) parts with the former due to particles insensitive to kinetics of the confining length $L$ and the latter due to particles moving synchronously with the kinetics of $L$.
From Eq.(\ref{force}) we see:
\begin{eqnarray}
\label{5.15}
\bar{F}L^3-\frac{\hbar^2}{2m}N^2\left[1+\frac{4\pi^2}{3}L_{0}^2\left(\frac{mkT}{\hbar^2}\right)^2 \left(\frac{N}{L_{0}}\right)^{-2}+\cdots\right]\nonumber\\
=\frac{m}{8}N^2L^3\ddot{L}\left[1+\frac{4\pi^2}{3}L_{0}^2\left(\frac{mkT}{\hbar^2}\right)^2 \left(\frac{N}{L_{0}}\right)^{-2}+\cdots\right].\nonumber\\
\end{eqnarray}
which is the extension of 1D Poisson's adiabatic law ($FL^3=const.$) to the case of a rapid piston.

Similarly, using Eq.(\ref{hamil-exp}), the internal energy for the expanding cavity is calculated as
\begin{eqnarray}
\label{energy}\label{soft-int-low}
\bar{U}&=&{\rm{Tr}}(\rho \hat{H}_{FF})=2\sum_{n=0}^\infty E_n f_n=2\int_0^\infty E D(E)f(E)dE\nonumber\\
&=&N^2\left(\frac{\hbar^2}{4mL^2}-\frac{m}{8} L\ddot{L}+\frac{m}{8} \dot{L}^2\right)\nonumber\\ 
&\times&
\left[1+\frac{4\pi^2}{3}L_{0}^2\left(\frac{mkT}{\hbar^2}\right)^2 \left(\frac{N}{L_{0}}\right)^{-2}+\cdots\right].\nonumber\\
\end{eqnarray}
As in the case of the force,  the internal energy consists of two contributions with one coming from particles insensitive to the kinetics of $L$ and the other from particles synchronously moving with kinetics of $L$. 
Combining Eqs.(\ref{force}) and (\ref{energy}), we have
\begin{eqnarray}
\label{4.33}
\bar{F}L-2\bar{U}&=&\left(\frac{3m}{8}N^2 L\ddot{L}-\frac{m}{4}N^2\dot{L}^2\right)\nonumber\\
&\times&\left[1+\frac{4\pi^2}{3}L_{0}^2\left(\frac{mkT}{\hbar^2}\right)^2 \left(\frac{N}{L_{0}}\right)^{-2}+\cdots\right],\nonumber\\
\end{eqnarray}
which stands for a fast forward variant of the quantum analogue of  1D Bernoulli's formula ($FL=2U$). 

The right-hand  side of Eq.(\ref{5.15}) is proportional to $\ddot{L}$ and that of Eq.(\ref{4.33}) consists of terms proportional to $\ddot{L}$ and $\dot{L}^2$. These terms, which are non-perturbative and time-reversal symmetric, comes from particles synchronously moving with kinetics of the confining size $L$ as explained below Eqs.(\ref{force}) and (\ref{soft-int-low}).

\subsection{High-temperature $(T \gg T_{0})$ and low-density region}
We shall then investigate the opposite limit, i.e., the high-temperature and low-density quasi-classical regime.
Here we shall have recourse to a high-temperature expansion of Fermi-Dirac distribution 
with  $\mu<0$:
\begin{eqnarray}\label{f51}
f(E)\equiv \frac{1}{e^{\beta(E-\mu)}+1}=\sum_{n=1}^\infty (-1)^{n-1} e^{-n\beta (E-\mu)}.\nonumber\\
\end{eqnarray}
Here
\begin{eqnarray} 
e^{\beta\mu}=\frac{\hbar^2N}{2mL_{0}^2kT}\left(1+ \frac{\hbar^2N}{4mL_{0}^2kT}+\cdots\right).
\end{eqnarray}
With use of the above equations we can obtain the force
\begin{eqnarray}\label{force-soft-high}
\bar{F}=2\sum^{\infty}_{n=0}f_{n}F_{n}&=&N\frac{mL_0^2kT}{\hbar^2}\left(\frac{2\hbar^2}{mL^3}+\frac{1}{2}m\ddot{L}\right)
\nonumber\\
&\times&\left[1+\frac{N}{8L_{0}^2}\frac{\hbar^2}{mkT}+\cdots\right].\nonumber\\
\end{eqnarray}
Since the prefactor ($\frac{mL_0^2kT}{\hbar^2}$) is dimensionless, the leading term of the above force consists of the adiabatic ($\frac{2\hbar^2}{mL^3}$) and nonadiabtic ($\frac{1}{2}m\ddot{L}$) terms with the former coming from particles insensitive to the kinetics of $L$ and the latter from particles synchronously moving with the kinetics of $L$.
Then we see the extension of Poisson's adiabatic law to the case of a rapid piston:
\begin{eqnarray}\label{PO-FF-Hi}
\bar{F}L^3-2L_{0}^2NkT\left[1+\frac{N}{8L_{0}^2}\frac{\hbar^2}{mkT}+\cdots\right]\nonumber\\
=\frac{1}{2}\frac{m^2}{\hbar^2}L_{0}^2L^3\ddot{L}NkT
\left[1+\frac{N}{8L_{0}^2}\frac{\hbar^2}{mkT}+\cdots\right].\nonumber\\
\end{eqnarray}
Similarly, the internal energy is given by
\begin{eqnarray}\label{soft-int-high}
\bar{U}&=&N\frac{mL_0^2kT}{\hbar^2}\left( \frac{\hbar^2}{mL^2}-\frac{1}{2}mL\ddot{L}+m\dot{L}^2\right) \nonumber\\
&\times&\left[1+\frac{N}{8L_{0}^2}\frac{\hbar^2}{mkT}+\cdots\right].
\end{eqnarray}
As in the case of the force, the internal energy consists of two parts with one from particles insensitive to the kinetics of $L$ and the other from particles synchronously moving with the kinetics of $L$.
The fast forward variant of Bernoulli's formula in the quasi-classical region is given by
\begin{eqnarray}\label{BB-FF-Hi}
\label{4.38}
\bar{F}L-2\bar{U}&=& \left(\frac{3}{2}\frac{m^2}{\hbar^2}L_{0}^2L\ddot{L}-2\frac{m^2}{\hbar^2}L_{0}^2\dot{L}^2\right)\nonumber\\
&\times&NkT\left[1+\frac{N}{8L_{0}^2}\frac{\hbar^2}{mkT}+\cdots\right].
\end{eqnarray}
The right-hand  sides of Eqs.(\ref{PO-FF-Hi}) and(\ref{BB-FF-Hi}) include  the nonadiabatic terms proportional to $\ddot{L}$ and $\dot{L}^2$, which come from particles synchronously moving with kinetics of the confining length $L$, 
as explained below Eqs.(\ref{force-soft-high}) and (\ref{soft-int-high}).

In closing this Section, we should note: During the fast-forward time region ($0\le t \le T_{FF}$), there is no transition among different quantum states, namely, the fast-forward dynamics is the population $(\rho_{nn})$-preserving cooling or heating process. Let's assume the ideal gas to have the equilibrium temperature $T(0)$ and effective temperature $T(t)$ at the initial $(t=0)$ and fast-forward time 
$(t>0)$, respectively. Then the $n$-th level population at  both temperatures should be identical, if there is the equality, 
 \begin{eqnarray}\label{rapid-temp}
 f\left(E_{n}(L(\Lambda(t=0))), T(0)\right)=f\left(E_{n}(L(\Lambda(t))), T(t)\right),\nonumber\\
\end{eqnarray}
 in the Fermi-Dirac distribution. In other words, Eq.(\ref{rapid-temp}) defines the effective temperature $T(t)$ during the fast-forward time. Noting that the Fermi-Dirac distribution is a function of $\frac{E}{kT}$, Eq.(\ref{rapid-temp}) is satisfied by imposing the condition
 \begin{eqnarray}
 \frac{\hbar\omega_0}{kT(0)}=\frac{\hbar\omega(L(\Lambda(t)))}{kT(t)}
 \end{eqnarray}
 or, with use of Eq.(\ref{FF-paramet}),
 \begin{eqnarray}\label{eff-temp}
 T(0)L_0^2=T(t)L^2(\Lambda(t)).
\end{eqnarray}

In all equations in this Section, we so far took $T=T(0)$. From now on, whenever we see $TL_{0}^2$, it will be read as $TL^2$ where $T(=T(t))$ and $L(=L(\Lambda(t)))$ are values in the fast-forward  time region $(0\le t \le T_{FF})$. Table 1 shows the fast-forward variants of Bernoulli's formula and Poisson's adiabatic equations obtained in this Section, where $\bar{F}$, $L$ and $T$ are are not quasi-static variables but rapidly-changing state variables at the same time $t$ $(0\le t \le T_{FF})$.

In the fast forward variants of equation of states, the nonadiabatic (NAD) contribution overwhelms the adiabatic one, 
if $mL\ddot{L}, m\dot{L}^2  \gg  \frac{\hbar^2}{mL^2}$, namely, if the energy of particles synchronously moving with kinetics of the confining size ($L$) is much larger than that of bulk particles which is insensitive to the kinetics of $L$. In the very rapid piston, therefore,  the feature of the fast forward variants of Bernoulli's formula and Poisson's adiabatic equation is completely different from that of the original equilibrium versions.

\section{The case of hard-walled confinement}\label{sec6}
We now investigate 1D quantum box with a moving wall. The dynamics of a particle is governed by 
\begin{eqnarray}\label{PH}
i\hbar\frac{\partial\psi}{\partial t}=H_{0}\psi=-\frac{\hbar^2}{2m}\partial_{x}^{2}\psi
\end{eqnarray}
with the time-dependent box boundary conditions as $\psi(x=0,t)=0$ and $\psi(x=L(t),t)=0$. 
$L(t)$ is assumed to change adiabatically as $L(t)=L_{0}+\epsilon t$.

The adiabatic eigenvalue problem related to Eq.(\ref{PH}) gives eigenvalues and eigenstates as
\begin{eqnarray}\label{EP}
E_{n}&=&\frac{\hbar^2}{2m}\left( \frac{\pi n}{L}\right)^2,\nonumber\\
\phi_{n}&=&\sqrt{\frac{2}{L}}\sin\left( \frac{\pi n}{L}x\right) . 
\end{eqnarray}
The phase $\theta$ which the regularized state acquires is given using the formula in Eq.(\ref{2.11}), as 
\begin{eqnarray}\label{theta}
\partial_{x}\theta&=&-\frac{m}{\hbar}\frac{1}{\phi_{n}^{2}}\partial_{L}\int_{0}^{x}\phi_{n}^{2}\mathrm{d}x=\frac{m}{\hbar}\frac{x}{L},\nonumber\\\theta&=&\frac{m}{2\hbar}\frac{x^2}{L}.
\end{eqnarray}
Thanks to the real nature of $\phi_{n}$, we find $\eta=0$ in Eq.(\ref{2.10}) and see that $\tilde{V}$ is vanishing.

Applying the fast forward scheme in Section \ref{sec2} and taking the asymptotic limit  ( $\epsilon\rightarrow 0, \bar{\alpha}\rightarrow \infty$ with $\epsilon\alpha=v(t)$ ), the fast forward state becomes
\begin{eqnarray}\label{psi-ff}
\psi_{FF}=\phi_{n}\left(x,L(\Lambda(t))\right)e^{i\frac{m\dot{L}}{2\hbar L}x^2} e^{-i\frac{\hbar}{2m}\left(\pi n \right)^2\int_{0}^{t}\frac{\mathrm{d}t'}{L^2(\Lambda(t'))} }, \nonumber\\ 
\end{eqnarray}
where $L\left( \Lambda(t)\right) =L_{0}+\int_{0}^{t}v(t')\mathrm{d}t'$
with the time scaling factor $v(t)$ given by $v(t)=\bar{v}\left( 1-\cos \frac{2\pi}{T_{FF}}t\right)$.
From Eq.(\ref{3.14}),  the fast forward potential is given by
\begin{eqnarray}
\label{6.6}
V_{FF}=-\frac{m}{2}\frac{\ddot{L}}{L}x^2,
\end{eqnarray}
which agrees with the existing references \cite{mako,cam2,jar2}. 
 
The TDSE for the fast-forward state is written as 
$i\hbar \frac{\partial }{\partial t}\psi_{FF}(x,t) =H_{FF}\psi_{FF}(x,t)$
with 
\begin{equation}
H_{FF}=-\frac{\hbar^2}{2m}\partial_x^2 +V_{FF}.
\end{equation}

As in the previous Section, the force operator is  now given by 
\begin{eqnarray}
\hat{F}=\frac{\hat{p}^2}{mL}-\frac{\dot{L}}{2L^2}\left(\hat{x}\hat{p}+\hat{p}\hat{x}\right)+\frac{m\ddot{L}}{2L^2}\hat{x}^2.
\end{eqnarray}

We now proceed to the statistical treatment of noninteracting particles in the case of the hard-walled confinement.  As in Eq.(\ref{rho-expand}), the solution of the von Neumann equation is:
\begin{eqnarray}
\rho=\sum^{\infty}_{n=1}\left|n\right\rangle f_{n}\left\langle n\right|.
\end{eqnarray}
 $f_{n}$ is the Fermi-Dirac distribution at $t=0$, i.e., $f_{n}=\frac{1}{e^{\beta\left(E_{n}(L(\Lambda(t=0)))-\mu\right)}+1}$ with $\beta=\beta(0)$. This $f_{n}$ can be replaced by $f_{n}=\frac{1}{e^{\beta(t)(E_{n}(L(\Lambda(t)))-\mu)}+1}$ because of the population-preserving nature of the fast-forward of adiabatic dynamics as described at the end of the previous Section. 
The time independence of $f_n$ is equivalent to introducing the equality $\beta(t)E_n(L(\Lambda(t)))=\beta(0)E_{n,0}$, which reduces to Eq.(\ref{eff-temp}) with use of Eq.(\ref{EP}).
 Below we shall take $\beta(\equiv \frac{1}{kT})$, $L$ and $E$ are time-dependent variables in the fast-forward time range $0\le t \le T_{FF}$.
The relevant matrix elements of force and Hamiltonian are as follows:
\begin{eqnarray}
F_{n}&=&\left\langle n\right|\hat{F} \left|n\right\rangle\nonumber\\
&=&\frac{\hbar^2\pi^2n^2}{mL^3}+\left(\frac{1}{6}-\frac{1}{4\pi^2n^2}\right)m\ddot{L},
\end{eqnarray}
\begin{eqnarray}
 E_{n}&=&\left\langle n\right|H_{FF} \left|n\right\rangle\nonumber\\
&=&\frac{\hbar^2\pi^2n^2}{2mL^2}+\left(\frac{1}{6}-\frac{1}{4\pi^2n^2}\right)(m\dot{L}^2-mL\ddot{L}).\nonumber\\
\end{eqnarray}
from which we confirme:$F_{n}=-\frac{\partial}{\partial L}E_{n}$.
Density of states $D(E)$ is given by
\begin{eqnarray}
D(E)=\sqrt{\frac{m}{2}}\frac{L}{\hbar\pi}E^{-1/2}.\nonumber\\
\end{eqnarray}

Let us calculate the statistical mean of force $\bar{F}$ and internal energy $\bar{U}$ .
Since the way of calculation is the same as in the previous Section, we shall show only the calculated results in the following.

\subsection{Low-temperature $(T\ll T_0)$ and high-density region}

Firstly, we must find the chemical potential
\begin{eqnarray}
N&=&2\int\limits_{0}^{\infty}D(E)f(E)dE \nonumber\\
&=&\frac{2\sqrt{2m}L}{\pi \hbar}\mu^{1/2}\Biggl(1-\frac{\pi^2}{24}(kT)^2\mu^{-2} +\cdots\Biggr),\nonumber\\
\end{eqnarray}
from which we obtain the chemical potential
\begin{eqnarray}
\mu=\frac{\pi^2\hbar^2}{8m}\Biggl(\frac{N}{L} \Biggr)^2\Biggl(1+\frac{16}{3\pi^2}\left(\frac{mkT}{\hbar^2}\right)^2\Biggl(\frac{N}{L}\Biggr)^{-4}+\cdots \Biggr).\nonumber\\
\end{eqnarray}

At low-temperature ($T\ll T_0$), the expectation of force is
\begin{eqnarray}\label{HD-f-low}
\bar F&=&\frac{\pi^2 \hbar^2}{12m}\frac{N^3}{L^3}\Biggl[1+\frac{24}{\pi^2}\left(\frac{mkT}{\hbar^2}\right)^2\Biggl(\frac{N}{L}\Biggr)^{-4}+\cdots\Biggr]\nonumber\\
&+&\frac{N}{6}m\ddot{L} \nonumber\\
&\times&\Biggl[1+\frac{6}{\pi^2}\frac{1}{N^2}\left(1+\frac{16}{3\pi^2}\left(\frac{mkT}{\hbar^2}\right)^2\Biggl(\frac{N}{L}\Biggr)^{-4}+\cdots \right)\Biggr]. \nonumber\\
\end{eqnarray}
The force consists of the adiabatic (1st line) and nonadiabatic (2nd line) contributions with the former coming from bulk particles insensitive to kinetics of the wall and the latter from near-wall particles synchronously moving with the wall motion. While the contribution from bulk particles is of $O(N^3)$ and that of near-wall particles is of $O(N)$, which is in marked contrast with the soft-wall confinement in the previous Section where both kind of contributions  are commonly of $O(N^2)$. 

Internal energy for our system is given by
\begin{eqnarray}\label{HD-e-low}
\bar U&=&\frac{\pi^2 \hbar^2}{24m}\frac{N^3}{L^2}\Biggl[1+\frac{24}{\pi^2}\left(\frac{mkT}{\hbar^2}\right)^2\Biggl(\frac{N}{L}\Biggr)^{-4}+\cdots\Biggr]\nonumber\\
&-&\frac{N}{6}(mL\ddot{L}-m\dot{L}^2) \nonumber\\
&\times&\Biggl[1+\frac{6}{\pi^2}\frac{1}{N^2}\left(1+\frac{16}{3\pi^2}\left(\frac{mkT}{\hbar^2}\right)^2\Biggl(\frac{N}{L}\Biggr)^{-4}+\cdots \right)\Biggr],
 \nonumber\\
\end{eqnarray}
which consists of the term of $O(N^3)$ from bulk particles insensitive to the wall motion and the one of $O(N)$ from near-wall particles synchronously moving with the wall motion.
Combining Eq.(\ref{HD-f-low}) with Eq.(\ref{HD-e-low}), we can obtain the fast-forward variants of Poisson's adiabatic equation and Bernoulli's formula in the low-temperature and high density region, which are
given in Table 1. Here the nonadiabatic contribution which comes from near-wall particles is by the factor of $O(N^{-2})$ less than the adiabatic one from bulk particles. The role of near-wall particles moving synchronously with the moving boundary was also pointed out in the context of quantum fluctuation theorem \cite{TM-qf,QJ-qf}.

\subsection{High-temperature $(T \gg T_{0})$ and low-density region}
Here we obtain the chemical potential by using 
\begin{eqnarray}
e^{\beta\mu}=\frac{N}{L}\sqrt{\frac{\pi\hbar^2}{2mkT}}\left(1+\frac{N}{2L}\sqrt{\frac{\pi\hbar^2}{mkT}}+\cdots\right).\nonumber\\
\end{eqnarray}
Now, we will calculate $\bar{F}$ for semiclassical region i.e. high-temperature and low-density region
\begin{eqnarray}
\label{6.21}
\bar F&=&\frac{NkT}{L}\left[1+\frac{N}{4L}\sqrt{\frac{\pi\hbar^2}{mkT}}+\cdots\right] \nonumber\\
&+&\frac{N}{6}m\ddot{L}\Bigl(1+\frac{3\pi\hbar^2}{2mL^2 kT}\Bigr).
\end{eqnarray}

Similarly the internal energy is given by
\begin{eqnarray}
\label{6.23}
\bar U&=&\frac{1}{2}NkT\left[1+\frac{N}{4L}\sqrt{\frac{\pi\hbar^2}{mkT}}+\cdots\right] \nonumber\\
&-&\frac{N}{6}(mL\ddot{L}-m\dot{L}^2) \Bigl(1+\frac{3\pi\hbar^2}{2mL^2 kT}\Bigr).\nonumber\\
\end{eqnarray}
In both the force and internal energy, the adiabatic contribution coming from bulk particles and nonadiabatic one from near-wall particles are commonly of $O(N)$, which differs from the characteristics in the low-temperature region of hard-wall confinement.
The above issues are reflected on the fast-forward variants of Poisson's adiabatic equation and Bernoulli's formula, which are
given in Table 1.

In the hard wall confinement, the Poisson's adiabatic equation has a new term proportional to $\ddot{L}$, and the Bernoulli's formula includes  two terms proportional to $\dot{L}^2$ and to $\ddot{L}$. The state variables $\bar{F}$, $L$ and $T$ are not quasi-static, but rapidly-changing variables during the fast-forward time range $(0\le t \le T_{FF})$.  These discoveries  are the same as in the case of soft-wall confinement. However, the criteria that the nonadiabatic (NAD) contribution overwhelms the adiabatic one is more subtle: The NAD contribution dominates the equation of states  in the low-temperature and high density region, if the energy of  near-wall particles ($mL\ddot{L}, m\dot{L}^2$) is much larger than that of bulk 
particles ($\frac{\hbar^2}{mL^2}$) multiplied by $N^2$,  and in the high-temperature and low-density region, if the energy of near-wall particles is much larger than $kT$ (: classical kinetic energy).

\begin{widetext}
{\bf Table 1.} Equation of states in thermally-isolated isentropic process for Fermi gas in the fast-forward protocol. 
$L, \bar{F}$ and $T$ are the cavity size, statistical mean of force and effective temperature defined 
by Eq.(\ref{FF-paramet}), Eq.(\ref{st-mean-F}) and Eq.(\ref{eff-temp}), respectively.

\begin{center}
\begin{tabular}{|c|l|l|} \hline
Equation of states\\
& low-temperature quantal region & high-temperature quasi-classical region\\ \hline  
\hline
Poisson's adiabatic\\ equations\\ 
soft-wall confinement&
$\bar{F}L^3-\frac{\hbar^2}{2m}N^2
\left[1+\frac{4\pi^2}{3}L^2\left(\frac{mkT}{\hbar^2}\right)^2 (\frac{N}{L})^{-2}+\cdots\right]$
 &
$\bar{F}L^3-2L^2NkT
\left[1+\frac{N}{8L^2}\frac{\hbar^2}{mkT}+\cdots\right]$\\ 
 &
$= \frac{m}{8}N^2 L^3\ddot{L}
\left[1+\frac{4\pi^2}{3}L^2\left(\frac{mkT}{\hbar^2}\right)^2 (\frac{N}{L})^{-2}+\cdots\right]$ 
&
$=\frac{1}{2}\frac{m^2}{\hbar^2}L^5\ddot{L}NkT
\left[1+\frac{N}{8L^2}\frac{\hbar^2}{mkT}+\cdots\right]$\\

hard-wall confinement&
$\bar{F}L^3-\frac{\pi^2 \hbar^2}{12m}N^3\Biggl[1+\frac{24}{\pi^2}(\frac{mkT}{\hbar^2})^2(\frac{N}{L})^{-4}+\cdots\Biggr]$
 &
$\bar{F}L^3-L^2NkT\left[1+\frac{N}{4L}\sqrt{\frac{\pi\hbar^2}{mkT}}+\cdots\right]
$\\ 
 &
$=\frac{N}{6}mL^3\ddot{L}\Biggl[1+\frac{6}{\pi^2}\frac{1}{N^2}\left(1+\frac{16}{3\pi^2}\left(\frac{mkT}{\hbar^2}\right)^2(\frac{N}{L})^{-4}+\cdots \right)\Biggr]$ 
&
$=\frac{N}{6}mL^3\ddot{L}\Bigl(1+\frac{3\pi\hbar^2}{2mL^2 kT}\Bigr)$\\
\hline
Bernoulli's formula\\
soft-wall confinement &
$\bar{F}L-2\bar{U}=N^2\left(\frac{3m}{8} L\ddot{L}-\frac{m}{4}\dot{L}^2\right)$
 &
$\bar{F}L-2\bar{U}=\left(\frac{3}{2}\frac{m^2}{\hbar^2}L^3\ddot{L}-2\frac{m^2}{\hbar^2}L^2\dot{L}^2\right)NkT$\\ 
 &
$\times\left[1+\frac{4\pi^2}{3}L^2\left(\frac{mkT}{\hbar^2}\right)^2 (\frac{N}{L})^{-2}+\cdots\right]$ 
&
$\times\left[1+\frac{N}{8L^2}\frac{\hbar^2}{mkT}+\cdots\right]$\\

hard-wall confinement &
$\bar F L-2\bar{U}$
 &
$\bar{F}L-2\bar{U}$\\ 
&
$ =N(\frac{1}{2}mL\ddot{L}-\frac{1}{3}m\dot{L}^2)$
&
$=N(\frac{1}{2}mL\ddot{L} -\frac{1}{3}m\dot{L}^2) $\\
 &
$\times \Biggl[1+\frac{6}{\pi^2}\frac{1}{N^2}\left(1+\frac{16}{3\pi^2}\left(\frac{mkT}{ \hbar^2}\right)^2(\frac{N}{L})^{-4}+\cdots \right)\Biggr]$ 
&
$\times \Bigl(1+\frac{3\pi\hbar^2}{2mL^2 kT}\Bigr)$\\
\hline

\end{tabular}

\end{center}
\end{widetext}

\section{Summary and discussions}\label{sec7}

Applying the idea of fast forward of adiabatic control of 1D confined systems, we investigated the nonequilibrium equation of states of an ideal quantum gas (Fermi gas) confined to  rapidly dilating soft-wall and hard-wall cavities. We showed the fast-forward variants of Poisson's adiabatic equation and Bernoulli's formula  which bridges the force and internal energy.
Confining ourselves to the thermally-isolated isentropic process and using the exact solution of the von Neumann equation, statistical means of the adiabatic and non-adiabatic (time-reversal symmetric) forces are evaluated in both the low-temperature quantum-mechanical  and high temperature quasi-classical regimes. 

Reflecting the fact that the fast-forward dynamics is population-preserving cooling or heating process, the state variables such as the statistical mean of force ($\bar{F}$), cavity size ($L$) and effective temperature ($T$) are not quasi-static, but rapidly-changing variables.
We elucidated the non-adiabatic (NAD) contributions to Poisson's adiabatic equation
and to Bernoulli's formula.  While the adiabatic contribution comes from ideal bulk particles insensitive to kinetics of $L$, NAD one comes from particles synchronously moving with the kinetics of  $L$, and are proportional to the acceleration ($\ddot{L}$)  and square of the velocity ($\dot{L}$). The relative ratio of NAD and adiabatic contributions is independent from the particle number ($N$)  in the case of soft wall confinement, whereas such a ratio is controllable in the case of hard-wall confinement.  We also revealed the condition when NAD contribution overwhelms the adiabatic one and thoroughly changes the standard form of the equilibrium equation of states.

In the analysis of nanoscale heat engine based on Fermi gas, one further needs nonequilibrium equation of states in the fast forward of the isochore process where the heat transfer between the gas and thermal reservoir is far from quasi-static, which will be investigated as a next challenge.

{\em Acknowledgments.} We are grateful to Y. Musakhanov and M. Okuyama for several critical comments.

%


\begin{thebibliography}{99}
\bibitem{adiab} In this paper, we assign the terminology "adiabatic" to an extremely-slow quantum dynamics which prevents the transition among different quantum states, while "thermally-adiabatic" means "thermally-isolated".
\end{thebibliography}
\end{document}